\documentstyle[twocolumn,aps,epsfig]{revtex}
\begin{document}
\draft
\title{Posture Sway and the Transition Rate for a Fall}

\author{R.K. Koleva, A. Widom and  D. Garelick}
\address{Physics Department, Northeastern University, Boston MA 02115}
\author{Meredith Harris} 
\address{Department of Physical Therapy, Northeastern University, Boston 
MA 02115}
 
\maketitle

\begin{abstract}
Postural body sway displacements for quiet standing subjects 
(measured with a new ultrasonic device) are reported. Two of the 
well known strategies for balancing, namely ankle and hip movements 
were probed. The data is modeled using a Fokker-Plank-Langevin 
stochastic theory. Both analytic and computer simulation techniques 
are employed. The Kramers transition rate for a fall is expressed as 
a function of experimental parameters. The root mean square velocity 
is especially important in determining the fall probability.
\end{abstract}

\pacs{87.19.St, 87.50.Kk, 05.40.Ca}

\section{Introduction}

A problem of considerable interest in human body motion dynamics is the 
prediction of the probability to fall in a given environment\cite{3,32}. 
Essential information is contained in the small amplitude displacement 
noise exhibited by the body during quiet standing. Our purpose is to explore 
how the quiet standing noise data may be used to compute the probability 
(more precisely the transition rate per unit time) of  falling in various 
external environments.

In the previous work of other groups, the data on body displacement noise 
were taken using a multicomponent force plate upon which the subject 
quietly stands\cite{3,1}. In the present work, we employ a sound wave 
assessment (SWA) device which measures body displacements using sound 
wave echoes; e.g., by measuring the time taken for a sound wave pulse 
to travel from the source to the standing subject and/or the time taken 
for a sound wave pulse to travel from the standing subject to a receiver. 
The experimental details concerning the SWA device will be discussed 
in Sec.II.

Three fundamental movement strategies to maintain a standing balance 
are well known\cite{2}: (i) body movement from the ankles, (ii) 
body movement from the hips, and (iii) stepping motions. Stepping 
motions will not be considered any further in this work which 
concerns quiet standing. As described in Sec.III, we find very 
different frequency scales for hip and ankle motions. The ankle 
motions proceed more slowly than the hip motions by about a 
factor of ten. Another feature of the ankle motions, 
also discussed in Sec.III, concerns the stability angle with respect 
to the vertical direction beyond which a subject, (without bending or 
stepping) will fall\cite{10}. The falling instability will be modeled by 
an appropriate potential in Sec.IV. Further, the noise itself will be 
model using the Fokker-Planck stochastic equation\cite{31} with a noise 
temperature intimately related to the mean square velocity fluctuations 
of the standing subject. 

Apart from the analytic work on stochastic equations, we have 
performed computer simulations of quiet standing subjects. In Sec.V 
we compare the computer simulations with data taken on subjects using 
the SWA device. The agreement between theory and experiment appears 
satisfactory.

In Sec.VI we make use of the Kramers potential well ``escape'' 
formula\cite{30}. This determines the transition rate per unit time 
for a subject to fall. The Kramers result for the transition rate 
$\Gamma_K $ for a fall is given by 
\begin{equation}
\Gamma_K =\left({\omega_0 Q\over \pi \sqrt{2}}\right)
\exp\left(-\left({\omega_0 a\over 2V}\right)^2\right),
\end{equation}
where $\omega_0 $ is the frequency of the ankle oscillation mode, 
$Q$ is the quality factor for this mode, $a$ is the critical 
displacement amplitude beyond which the ankle mode is unstable, 
and $V$ is the root mean square velocity which is determined by the 
environmental noise temperature.

In the discussion Sec.VII, we note that the transition rate 
$\Gamma_K $ for a fall has an essential singularity $\Gamma_K \to 0$  
as the root mean square velocity $V\to 0$. This implies that $V$ 
is the most sensitive parameter which may be used to determine 
the probability of falling. The implications will 
be discussed in the concluding Sec.VII of this work. 

\section{The Sound Wave Assessment Device}

The SWA device employs two small ultrasonic transducers each of which can 
send or receive ultrasonic signals. The first is positioned on a stable 
laboratory stand. The second is attached with belt to the lower back of 
quietly standing subject. Each transducer sends sixteen ultrasonic pulses 
every $3.3\times 10^{-2}\ sec$, which are later detected by the opposite 
transducer. The distance between the two transducers can then be obtained 
from the time required for the pulses to travel from the sender to the 
receiver. 

A single transducer system has been used in industry, e.g. by Polaroid 
for photography applications, in order to determine the distance to an 
object. The single transducer is used to both emit and detect its 
own echo. Our use of {\em two transducers} improves the accuracy by 
having less air absorption as well as less spatial dispersion 
(fanning out) of the pulses. Our position measurements are quiet precise; 
i.e. displacements are measured to within an instrumental accuracy of 
$0.02\ cm$ in a bandwidth of $12\ kHz$. This yields a displacement noise 
error of $\delta x\approx 2\ (\mu m/\sqrt{Hz})$. Thus, we can record the 
fine movements of a subject (both towards and away from a static 
transducer) as a function of time.

\section{Frequency Scales for Hip and Ankle Strategies}

Two fundamental body movement strategies\cite{2} (to maintain a standing 
balance) are apparent in the SWA measurements discussed below. 
These are the quiet standing ankle movements and hip movements, 
which occur in widely different frequency bands. 
The hip motions take place on a time 
scale of $\tau_{hip}\sim 2\ sec$, while the ankle motions take place 
on a time scale of $\tau_{ankle}\sim 30\ sec$. In order to measure 
ankle motions, the durations of our measurements were rather long, i.e. 
$3\ min$, while the shorter runs of $1\ min$ duration were sufficient 
to study hip motions. The short times scales, of less than $20\ sec$ 
were sufficient to verify the fractal diffusion estimates of hip motions 
which have appeared in the pioneering studies\cite{1,60,61,63} using the 
force plate technology.

\begin{figure}[htbp]
\begin{center}
\mbox{\epsfig{file=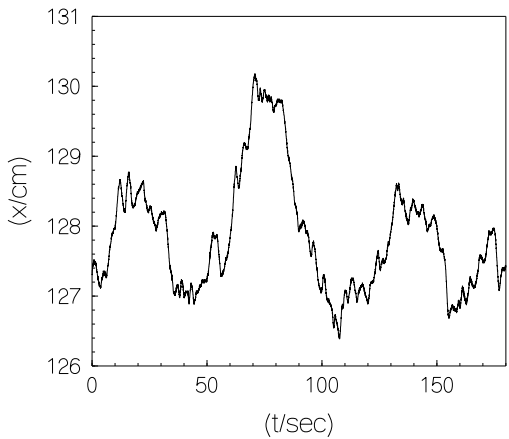,height=80mm}}
\caption{SWA measurement of $x(t)$ for a quiet standing subject. The large 
broad peaks and valleys are from ankle motions, and the smaller more rapid 
oscillations are from hip motions.}
\label{fig1}
\end{center}
\end{figure}

To illustrate the decomposition of movements into ankle motions and 
hip motions we plot, in FIG.1, the displacement as a function of time for a 
typical quiet standing subject. One notes the slowly varying drifts back 
and forth characteristic of ankle motions along with the more rapid 
(and in this case) much smaller hip motions.

Were the quiet standing subject to move without hip motions, i.e. move 
rigidly, then the stiff body would be held steady, if and only if the angle 
$\vartheta $ formed with the vertical were less then some critical angle 
$\vartheta_c $. The geometrical cone $\vartheta <\vartheta_c $ is called 
the sway envelop or the cone of stability\cite{10}. Typical values for the 
critical (forward) angle is $\vartheta_c \approx 8^o$\cite{4}.
For angles $\vartheta > \vartheta_c $, the ankles can no longer hold the 
subject upright. In the absence of a step or other supports 
(such as arm motions) a fall will take place.   
If one considers the displacement of the part of the body at a height 
$h$ above ground, then $a=h\tan \vartheta_c $ represents the critical 
displacement amplitude beyond which the ankle mode is unstable. Such 
critical displacements are normally measured using standard tests of the 
subjects forward or backward reach\cite{33,34}.

\section{Fokker-Planck-Langevin Theory}  

The Fokker-Planck-Langevin stochastic approach\cite{31} to 
quiet standing sway processes has been shown to be extremely 
useful\cite{40,50}. In what follows we apply this method to the 
ankle movements. As discussed in Sec.III, the ankle movements 
can be described by a slow oscillatory motion modulated by 
faster random noise. 

To mathematically describe the cone of stability, we model the 
{\em metastable} potential by the potential 
\begin{equation}
U(x)=-\left({m\omega_0^2a^2\over 4}\right)
\left[1-\left({x\over a}\right)^2\right]^2,
\end{equation} 
where $\omega_0$ is the frequency of the ankle oscillation mode, 
$m$ is the mass of the mode described by the displacement $x$, 
and $a$ is the critical displacement. 

The ankle mode, as shown in the experimental FIG.1, is far from a 
perfect oscillation. There exists a random force $f(t)$ and a finite 
quality factor $Q$ for the mode which, in the Langevin theory, are 
intimately related. The equation of motion for the mode may be 
written as 
\begin{equation}
m\left[\ddot{x}+\left({\omega_0\over Q}\right)\dot{x}+\omega_0^2x \right]
+U^\prime (x)=f(t)+F(t),
\end{equation}
where $F(t)$ is an applied force, and $f(t)$ is a ``white noise'' random 
force obeying 
\begin{equation}
\left<f(t)f(t^\prime )\right>_{noise}
=\left({2m\omega_0 k_BT_n\over Q}\right)
\delta (t-t^\prime ),
\end{equation} 
and where $T_n$ is noise temperature. The noise temperature is by no means 
equal to the ambient temperature of the room in which the subject stands. 
Rather the noise temperature describes all of those fluctuations which 
couple into the coordinate $x$. In particular, all of the environmental 
and internal fluctuations which couple into $x$ contribute to the  
root mean square velocity $V$ 
\begin{equation}
V^2=\left<\dot{x}^2\right> 
\end{equation}
which can be obtained from experimental data such as that pictured 
in FIG.1. The noise temperature $T_n$ is here {\em defined} in terms of 
the root mean squared velocity $V$ via the equipartition theorem\cite{62} 
\begin{equation}
\left({mV^2\over 2}\right)=\left({k_BT_n\over 2}\right).
\end{equation}
Thus, for the case in which the applied force is written $F(t)=m\beta (t)$, 
the random force is written $f(t)=m\alpha (t)$ and the potential 
$U(x)=m\phi (x)$, the Langevin equation reads 
\begin{equation}
\ddot{x}+\left({\omega_0\over Q}\right)\dot{x}+\omega_0^2x
+\phi^\prime (x)=\alpha (t)+\beta (t),
\end{equation}
where 
\begin{equation}
\phi (x)=-\left({\omega_0^2a^2\over 4}\right)
\left[1-\left({x\over a}\right)^2\right]^2,
\end{equation}
and 
\begin{equation}
\left<\alpha (t)\alpha (t^\prime )\right>_{noise}
=\left({2\omega_0 V^2\over Q}\right)
\delta (t-t^\prime ).
\end{equation}
Eqs.(7), (8) and (9), with a hip modulation given by 
\begin{equation}
\beta (t)=\beta_{max}cos(\omega_{hip}t)
\end{equation}
can be used to perform a computer simulation of the ankle mode of 
motion. 

\begin{figure}[htbp]
\begin{center}
\mbox{\epsfig{file=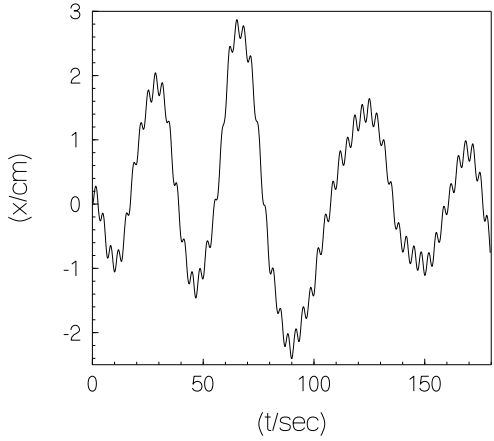,height=80mm}}
\caption{Simulated quiet standing posture sway with a frequency 
$\omega_0=0.125\ sec^{-1}$, a critical reach of $a=12\ cm$, a noise 
temperature corresponding to $V=0.22\ cm/sec$, a quality factor of 
$Q=20$, a hip modulation frequency of 
$\omega_{hip}=1.5\ sec^{-1}$, and a hip modulation amplitude of 
$\beta_{max}=0.37\ cm/sec^2$.}
\label{fig2}
\end{center}
\end{figure}
For the healthy quiet standing subject, shown in FIG.1, 
an output of the random motion simulation is shown in FIG.2.
The qualitative similarity between a number of quiet standing subjects 
and simulations allow us to conclude that the Langevin theory of 
postural sway is reasonable. 

The Langevin theory can also be analytically expressed as an equation 
for the probability $P(x,v,t)dx dv$ for the subject to have a velocity 
in the interval $dv$ and a displacement in the interval $dx$. For 
example, without the hip modulation, the Fokker-Planck equation reads 
$$
\left({\partial P\over \partial t}\right)+
v\left({\partial P\over \partial x}\right)=
$$
\begin{equation}
{\partial \over \partial v}
\left\{\left({\omega_0 v \over Q}\right)P+\phi^\prime (x)P \right\}
+\left({\omega_0 V^2 \over Q}\right)
\left({\partial ^2 P\over \partial v^2}\right).
\end{equation}
The Fokker-Planck and the Langevin formulations of the problem are 
equivalent. The former is useful for analytical calculations while the 
later is useful for computer simulations. 

\section{Postural Sway Data}

Not all of the quiet standing subjects measured exhibited large 
undamped ankle mode oscillations. For some quiet standing subjects 
the ankle movements were strongly damped. Shown in FIG.3 is an 
example of a subject with suppressed ankle movements. The hip 
oscillations are still clearly visible.

\begin{figure}[htbp]
\begin{center}
\mbox{\epsfig{file=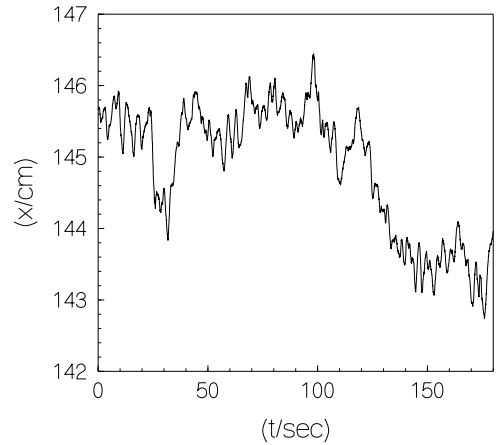,height=80mm}}
\caption{Shown is a quiet standing subject with a strongly damped 
ankle mode. The hip oscillations about the diffusive ankle motions are 
clearly visible.}
\label{fig3}
\end{center}
\end{figure}

The computer simulations for the overdamped case proceed exactly as 
previously described. However the quality factor of the mode is taken 
to be much smaller than for the underdamped case. 
For the quiet standing subject, shown in FIG.3, an output of 
the random motion simulation is shown in FIG.4. We again 
conclude, for the overdamped case, that the Langevin theory of 
postural sway is reasonable.

\begin{figure}[htbp]
\begin{center}
\mbox{\epsfig{file=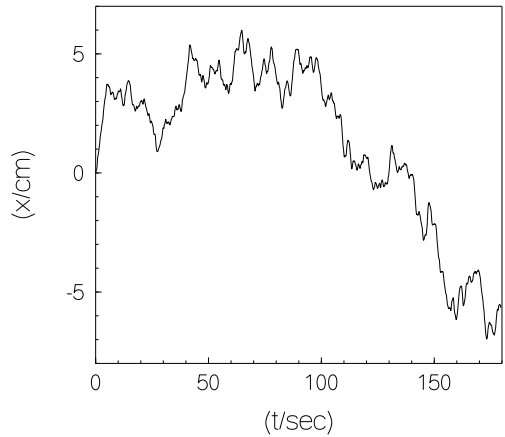,height=80mm}}
\caption{Simulated overdamped quiet standing posture sway with a 
frequency $\omega_0=0.125\ sec^{-1}$, a critical reach of $a=12\ cm$, a noise 
temperature corresponding to $V=0.31\ cm/sec$, a quality factor of 
$Q=0.1$, a hip modulation frequency of 
$\omega_{hip}=1.5\ sec^{-1}$, and a hip modulation amplitude of 
$\beta_{max}=0.56\ cm/sec^2$.}
\label{fig4}
\end{center}
\end{figure}

The ankle movement is more clearly seen in the mean square displacement 
fluctuation defined in mathematical Brownian motion theory as  
$$
\left<\Delta x(\tau )^2\right>=
$$
\begin{equation}
\lim_{T\to \infty}\ {1\over T}\int_{t_0-(T/2)}^{t_0+(T/2)}
\left|x(t+\tau )-x(t)\right|^2 dt.
\end{equation}
In physical systems the time $T$ of a run is finite. In our case 
$T=180\ sec$. We can then plot $\left<\Delta x(\tau )^2\right>$ 
on the interval $0<\tau <140\ sec$. 

For the underdamped experimental data of FIG.1, we show in FIG.5 
a plot of $\left<\Delta x(\tau )^2\right>$. The ankle movements show 
an oscillation in a very clear form when the averaging procedure of 
Eq.(12) is performed. The hip movement modulations are largely smoothed 
away by this same averaging procedure. 

The data produced by the computer simulations can also be averaged 
according to Eq.(12) and then compared with mean square fluctuations 
taken from the experimental data. The hip movement modulations are 
barely noticeable in $\left<\Delta x(\tau )^2\right>$.  

\begin{figure}[htbp]
\begin{center}
\mbox{\epsfig{file=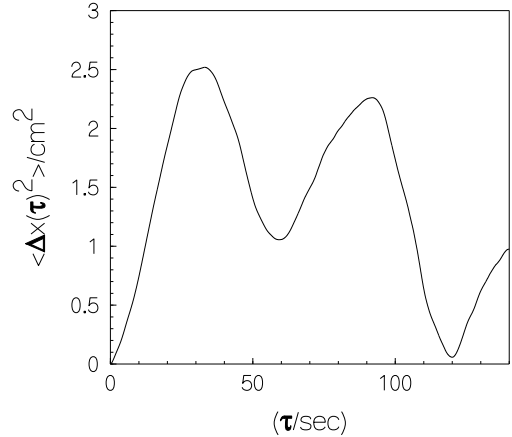,height=80mm}}
\caption{SWA measurement of $\left<\Delta x(\tau )^2\right>$ for a quiet 
standing subject taken from the raw data in FIG.1. The undamped oscillation 
of the ankle mode is clearly visible. The hip motions are smoothed out in 
the mean square displacement fluctuation.}
\label{fig5}
\end{center}
\end{figure}
In FIG.6 we have plotted $\left<\Delta x(\tau )^2\right>$ for the 
simulation in FIG.2. We note that the oscillations for underdamped 
ankle movements are clearly present although the amplitude of the 
oscillation are high when compared with the experimental FIG.5. 
However, we still conclude that the Langevin model is qualitatively 
reasonable.
\begin{figure}[htbp]
\begin{center}
\mbox{\epsfig{file=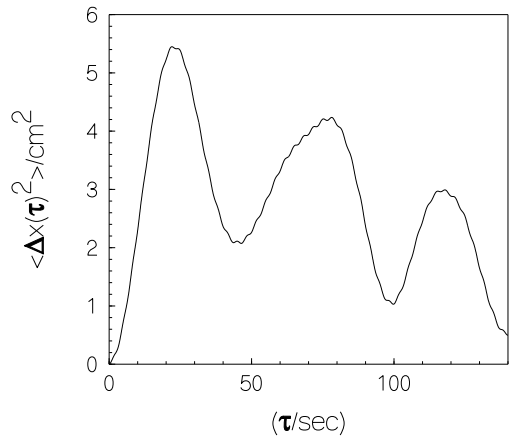,height=80mm}}
\caption{Simulated mean square fluctuation 
$\left<\Delta x(\tau )^2\right>$ with parameters the same as in FIG.2.}
\label{fig6}
\end{center}
\end{figure}

For the overdamped case, the data in FIG.3 gives rise to an experimental 
$\left<\Delta x(\tau )^2\right>$ shown in FIG.7. Notice for the 
overdamped case the absence of oscillations in the ankle motion. The 
behavior of $\left<\Delta x(\tau )^2\right>$ on the time scale shown 
is qualitatively not very far from simple Brownian diffusion. 

\begin{figure}[htbp]
\begin{center}
\mbox{\epsfig{file=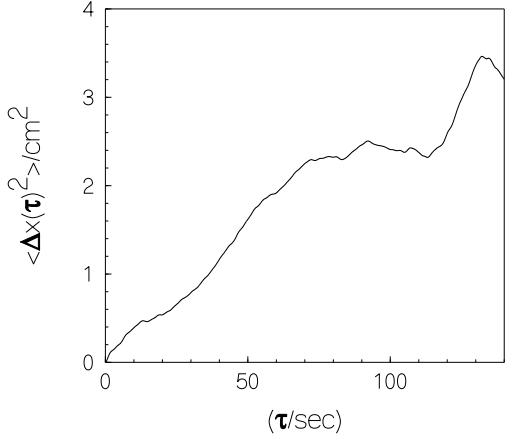,height=80mm}}
\caption{SWA measurement of $\left<\Delta x(\tau )^2\right>$ for a quiet 
standing subject taken from the raw data in FIG.3. The oscillation 
of the ankle mode overdamped.}
\label{fig7}
\end{center}
\end{figure}

This diffusion-like behavior is also present in the computer simulation 
shown in FIG.8. The simulated amplitudes are again large compared with 
the experimental amplitudes. 

\begin{figure}[htbp]
\begin{center}
\mbox{\epsfig{file=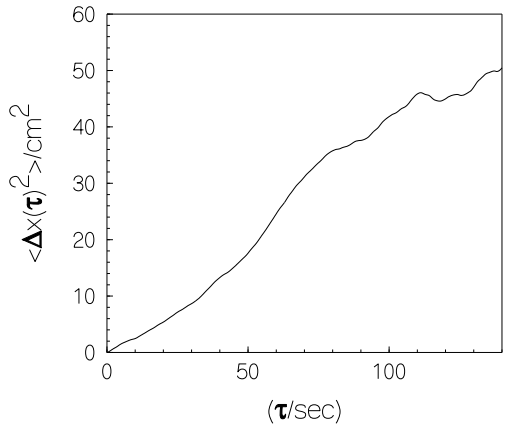,height=80mm}}
\caption{Simulated mean square fluctuation 
$\left<\Delta x(\tau )^2\right>$ with parameters the same as in FIG.4.}
\label{fig8}
\end{center}
\end{figure}

The Fokker-Plank-Langevin theory is mainly of qualitative significance 
when comparing simulations to experimental data.  

\section{Kramers ``Escape Rate'' for a Fall}

Although the form of the metastable potential in Eq.(2) is commonly used, 
the potential has not yet been shown to be unique. The qualitative features 
required are a minimum near $x=0$ and a barrier to falling near $x=a$.
Eq.(2) is plotted in FIG.9.
\begin{figure}[htbp]
\begin{center}
\mbox{\epsfig{file=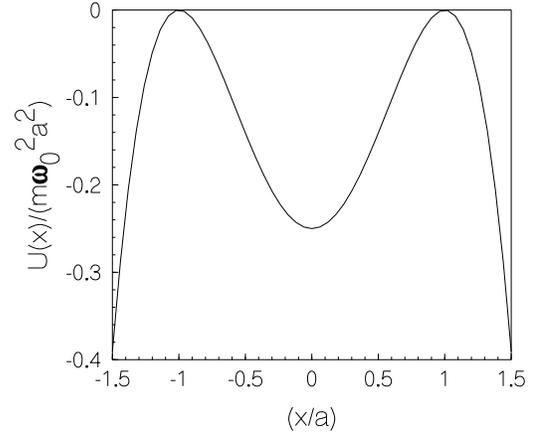,height=80mm}}
\caption{The model potential $U(x)$ for describing a fall.}
\label{fig9}
\end{center}
\end{figure}
A quiet standing subject exhibits oscillations about the potential minimum. 
The random force at a noise temperature $T_n$ will at some time push the 
displacement to values $|x|>a$ over a potential maxima at which time there 
is a fall. The potential barrier protecting the subject from a fall is 
given by 
\begin{equation}
\Delta U=U(|a|)-U(0)=\left({m\omega_0^2a^2\over 4}\right).
\end{equation}
Neglecting hip motion modulations, the Kramers equation\cite{30} for the 
escape rate contains the Boltzman factor for overcoming a barrier; It is 
\begin{equation}
\Gamma_K =\left({\omega_0 Q\over \pi \sqrt{2}}\right)
e^{-\Delta U/k_BT_n}\ .
\end{equation}
In terms of experimental parameters 
\begin{equation}
\Gamma_K =\left({\omega_0 Q\over \pi \sqrt{2}}\right)
\exp\left(-\left({\omega_0 a\over 2V}\right)^2\right),
\end{equation}

We used the parameters in Eq.(15) to describe twenty healthy subjects 
(between sixteen and forty five years of age) in street clothing with 
eyes wide open. The reach test length scale $a$ was calculated from 
the critical angle $\vartheta_c$ given in the literature\cite{4}. 
We find the Kramers predicted ``times for a fall'' to be anywhere 
from one week to three years. The time shortens considerably 
as the mean square velocity increases. Of course, this prediction 
by no means implies that these subjects will all experience a fall 
within the next three years! The fall transition rate of our model 
will be further discussed in the next and concluding section.

\section{Discussion}

We have exhibited measurements of displacement noise with 
quiet standing subjects along with computer simulations based on a 
Fokker-Plank-Langevin model. For all ranges of the quality factor 
$Q$, from underdamped $Q>>1$ to overdamped $Q<<1$, the Langevin equation 
computer simulations were in qualitative agreement with the data. 
In this regard, one should note that the same subject when measured 
on different days can exhibit both overdamped and/or underdamped behavior, 
as well a somewhat different effective noise temperatures $T_n$, 
i.e. a somewhat different root mean square velocity $V$. 

Since the Kramers transition rate for a fall depends 
sensitively on the root mean square velocity $V$, it follows that 
a persons transition rate for a fall can vary from day to day 
depending on the ``noise temperature''. For example, on a day when a 
person is tired we conjecture that a fall is more likely than on a day 
when the person is wide awake. However, we presently have no data with 
which to prove this conjecture.
  
While the above Fokker-Planck-Langevin theory seems to be qualitatively 
correct, the theory overestimates the displacement noise amplitudes and 
thereby also overestimates the transition rate for a fall. Apart from 
leaving out the step strategy for avoiding a fall, we believe that we 
have perhaps also underestimated the subtlety of the hip strategy. 
Thus far, the hip motions have been described by an added modulation 
force in our computer simulations, and this inclusion gives at least 
qualitative agreement with experimental data. Nevertheless, the hip 
modulations might be more strongly correlated with the center of mass 
body coordinate\cite{80,85} than is presently being included. Further 
improvements on the present model are presently being 
pursued on both the experimental and theoretical level.


\begin{thebibliography}{99}
\bibitem{3} T.E. Prieto, J.B. Myklebust, and B.M. Myklebust,
IEEE Trans. Rehab. Eng. {\bf 1}, 26 (1993).
\bibitem{32} B.E. Maki, P.J. Holliday, and G.R. Fernil, 
IEEE Trans. Biomed. Eng, {\bf 34}, 797 (1987). 
\bibitem{1} J.J. Collins and C.J. DeLuca, Phys. Rev. Lett. {\bf 73},
764 (1994). 
\bibitem{2} L.M. Nasher and G. McCollum, Behav. Brain. Sci. {\bf 8},
135 (1985). 
\bibitem{10} G. McCollum and T.K. Leen, J. Motor. Behav {\bf 21,}
225 (1989).
\bibitem{31} H. Risken, {\it The Fokker-Planck Equation: Methods of Solution  
and Applications} (Springer-Verlag, New York, 1996) 2nd ed. 
\bibitem{30} H.A. Kramers, Physica {\bf 7}, 284 (1940).
\bibitem{60} J.J. Collins, C.J. DeLuca,  A. Borrows, and L.A. Lipsitz,
Exp. Brain Res. {\bf 104,} 480 (1995).
\bibitem{61} S.L. Mitchell {\it et al}, Neurosci. Lett. {\bf 197},
133 (1995).
\bibitem{63} J.J. Collins and C.J. DeLuca, CHAOS {\bf 5}, 57 (1995).
\bibitem{4}{ L.M. Nasher, Sensory, neuromuscular and biomechanical
contributions to humal balance. In: Balance: Proceedings of the
American Physical Therapy Association Forum: Nashville, TN, 
June 13-15, (1989).}
\bibitem{33}{P.W. Duncam, D.K.Weiner, J. Chandler, and S. Studenski, 
J. Gerontol. {\bf 45,} M192 (1990).}
\bibitem{34}{L.D.B. Thornbahm and R.A. Newton, Phys. Ther. {\bf 76,}
576 (1996).}
\bibitem{40}{M. Lauk {\it et al}, Phys. Rev. Lett. {\bf 80}, 413 (1998).}
\bibitem{50}{C.C. Chow and J.J. Collins, Phys. Rev. {\bf 52,}, 
907 (1995).}
\bibitem{62}{L.E. Reichl, {\it A Modern Course in Statistical Physics,}
(Wiley, New York, 1998) 2nd ed.}
\bibitem{80} {S.H. Koozekanani, C.W. Stockwell, R.B. McGhee, and F. 
Firoosmand, IEEE Trans. Biomed. Eng. {\bf 27}, 605 (1980).}
\bibitem{85} {P.O. Riley, R.W. Mann, and W.A. Hodge, J. Biomech, 
{\bf 23}, 503 (1990).}

\end{thebibliography}
\end{document}